\documentclass[nologo,url,11pt,a4paper]{ETHpaper}
\usepackage{amssymb,amsmath} % Mathematical Symbols
\usepackage{subfig}
\usepackage{graphicx}
\usepackage[sort&compress]{natbib} % Bibliography
\usepackage[utf8]{inputenc}
\usepackage{color}
\pdfoutput=1
\begin{document}

\title{Quantifying the Impact of\\ Leveraging and Diversification on Systemic Risk}
\titlealternative{Quantifying the impact of leveraging and diversification on systemic risk}
%%Systemic Risk: Quantifying the impact of leveraging and diversification}
\author{Paolo Tasca$^\star$, Pavlin Mavrodiev, Frank Schweitzer}
\authoralternative{P. Tasca, P. Mavrodiev, F. Schweitzer}
\address{Chair of Systems Design, ETH Zurich, Weinbergstrasse 58, 8092
   Zurich, Switzerland}

\reference{Submitted \today }

\www{\url{http://www.sg.ethz.ch}}

\makeframing
\maketitle

\footnotetext{The authors
  acknowledge financial support from the ETH Competence Center
  ``Coping with Crises in Complex Socio-Economic Systems'' (CHIRP 1
  grant no.  CH1-01-08-2), the European FET Open Project
  ``FOC'' (grant no.  255987), and the Swiss National Science
  Foundation project ``OTC Derivatives and Systemic Risk in Financial
  Networks'' (grant no.  CR12I1-127000/1). ($\large{\star}$) Correspondence to Paolo
  Tasca, \texttt{ptasca@ethz.ch}.}
 
\begin{abstract}
  Excessive leverage, i.e. the abuse of debt financing, is considered one of the
  primary factors in the default of financial institutions. Systemic risk
  results from correlations between individual default probabilities
  that cannot be considered independent.  Based on the structural
  framework by Merton (1974), we discuss a model in which these
  correlations arise from overlaps in banks' portfolios. Portfolio
  diversification is used as a strategy to mitigate losses from
  investments in risky projects. We calculate an optimal level of
  diversification that has to be reached for a given level of excessive
  leverage to still mitigate an increase in systemic risk. In our
  model, this optimal diversification further depends on the market
  size and the market conditions (e.g. volatility). It allows to
  distinguish between a \textit{safe} regime, in which excessive leverage
  does \emph{not} result in an increase of systemic risk, and a
  \textit{risky} regime, in which excessive leverage cannot be
  mitigated leading to an increased systemic risk. Our results are of
  relevance for financial regulators.
% In a two-bank economy in which both banks
%     invest in a finite frictionless market of uncorrelated risky
%     projects under the Merton (1974) framework, we find that the
%     deterioration of their credit quality due to over-leveraging, does
%     not increase systemic risk if their
%     portfolios are diversified above a threshold $n^*$. Furthermore,
%     the market size increases substantially this threshold level.
\end{abstract}

 \vspace{0.2cm}

%\end{keyword}

\section{Introduction}

Systemic risk can be generally described as the risk that a failing
agent (e.g. a bank in a financial system or a firm in a supply system)
causes the failure of other agents such that failing cascades may
encompass the whole system
\citep{Battiston.Lorenz.ea2009FailureCascadesNetworks}. This approach
differs from other notions of risk \citep{embrechts2011modelling}
which treat the default of individual agents as an extreme event for
which the probability is calculated regardless of the interaction
among economic agents. Instead, a systemic notion of risk requires to
explicitly take into account two ingredients: (a) the stability
condition of individual agents, and (b) the impact of interactions
between agents on their mutual stability. In this paper, we apply this
systemic perspective to analyze the stability of a banking system,
dependent on the leverage and diversification of individual banks.

The stability of a single bank is commonly expressed by the
composition of its balance sheet and measured by the \emph{leverage},
e.g. by the debt-to-asset ratio. To quantify stability, we calculate
the individual default probability using the well established
structural framework by \citet{Merton1974PricingofCorporate} in which
the firm's leverage plays a major role.  This approach has become
popular among both academicians and practitioners thanks to its
tractability and simplicity. For a survey see \citep{bohn2000survey}.
Moreover, the Financial Stability \citet{board2010guidance} recommends
it as a building block in establishing a regulatory framework that can
cope with risk from systemic linkages.

These linkages arise from bilateral exposures (e.g. mutual claims) and
common assets which may lead to interlocking balance sheets
\citep[see][]{Kiyotaki.Moore2002Balance-SheetContagion}.  Including
these effects into a systemic notion of risk requires explicit
knowledge of banks' balance sheets and their change over time. This
information is mostly confidential, due to strategic issues, and
therefore unattainable for a system comprised of many banks. To avoid
the trap of insufficient data, we study the mutual impact of failing
banks by calculating their joint default probability (hereafter,
``systemic default probability'') based on their individual default
probabilities, as defined by the Merton framework, and their \emph{asset
correlation}. 
%The latter increases with the level of diversification. 
We assume that risk-averse banks hold diversified portfolios in order to mitigate their individual default risk. Therefore, in a
market with a finite number of (uncorrelated) investment opportunities, also called projects, asset correlation among banks
increases with diversification due to overlapping portfolios.  
%\textcolor{blue}{Indeed, overlapping portfolios is widely
%  considered to have been the primary vector of contagion in the recent 2007-2008 U.S. financial crisis. E.g.,
\cite{stein2009presidential} identifies similar portfolios, or ``crowding'', as a destabilizing mechanism in financial markets.

% Indeed, banks use diversified asset portfolios in order
% to mitigate their individual default risk. With a finite number of
% assets, this strategy is likely to result in overlapping portfolios
% among banks.

% So far, only few works \citep{cathcart2004multiple} have extended the structural framework of
% \citet{Merton1974PricingofCorporate} to calculate the default probability of many firms.

In our paper, we particularly focus on the impact of \textit{excessive leverage} and \textit{diversification} on the systemic
default probability and their possible interdependence. Excessive leverage refers to the practice of banks to engage in huge debt
to buy more assets, in order to increase their return on equity. As the
Financial Service Authority \citep{turner2009turner} and the
Financial Stability Board \citep{board2009report} jointly
point out, excessive leverage by banks increases systemic risk.  However, our hypothesis is that diversification can compensate
some of the hazards of excessive leverage, under optimal conditions. In our paper we identify two different regimes: (i) a
\emph{safe} regime in which an increase of leverage % , i.e. of risk to default
can be compensated by a proper diversification strategy, leading to a low systemic default probability, and (ii) a \emph{risky}
regime in which the same increase of leverage will lead to an increase of the systemic default probability, because the
diversification strategy is not adequate and/or the market conditions are adverse. The latter
consider, in our model, the size of the market, the market risk as measured by the volatility, and the time horizon.

% The increase of
% leverage beyond the ``normal'' level, a.k.a. \textit{excess leverage}, results from
% investments which are financed by huge debts that became attractive
% thanks to e.g. low interest rates.

The boundary between the two regimes is, in our model, given by a
critical level of diversification, $n^{\star}(f_{n},f_{a},N,\chi)$,
which depends on the market size $N$, the market conditions,  $\chi$
(which also include the volatility and the time horizon), and two
leverages, referring to a normal ($f_{n}$) and an abnormal ($f_{a}$)
level. The safe regime is given by $n\geq n^{\star}$, whereas the
risky regime is the opposite. We show that $n^{\star}$ increases with
$N$. This seems to be counter-intuitive because the larger the market,
the less should be the portfolio overlap for a given level of
diversification. However, this only holds if $n$ is above the critical
level of diversification. With increasing market size, banks have to
reach a larger level of diversification in order to be on the safe
side. Secondly, we show that excessive leverage
{expands} the risky regime and therefore the critical
level of diversification. Interestingly, there is not always a
combination of leverage and diversification to reach the safe
regime. Even with increasing market size, the market conditions may
prevent this, thus the systemic default probability is not reduced.

Our paper contributes to several research lines.  First, we show a
different way of how the well established
\citet{Merton1974PricingofCorporate} model can be extended to joint
defaults.  Second, we extend the previous literature on the optimal
diversification level pioneered by \cite{evans1968diversification} and  
\cite{Elton77}
 to
a systemic context. This allows to better understand the controversial
relation between the firm capital structure and diversification
\citep{Jensen76}. Finally, our work complements existing theoretical
literature that, after the 2007-2008 financial crisis, started to
investigate the role of risk diversification for the stability of the
financial system \citep{Brock2009Morehedginginstruments, Wagner,
  ibragimov2007limits,
  Battiston.Gatti.ea2009LiaisonsDangereusesIncreasing,
  stiglitz2010FullFinancial}. Our results about a critical level of
diversification can be used for improving macro-prudential
regulations, for example by enforcing a ceiling to excessive leverage.

\section{Model}
\label{Model}

\subsection{Banks and their portfolio}
\label{sec:banks}

We extend the \citet{Merton1974PricingofCorporate} framework to a set of $M$ banks such that for each bank $i \in M$ the following
balance-sheet identity holds true at any time $t$
\begin{equation}\label{eq:36}
  a_i(t)=h_i(T) + e_i(t) %\:,
\end{equation}
where $a_i(t)$ is the market value of bank $i$'s assets. $h_i(T)$ is the promised payment of its liabilities at maturity date
$T$.\footnote{The equation indirectly assumes only one seniority of bonds with the same maturity date $T$.} Finally, $e_i(t)$ is
the equity value which keeps the two sides of the balance sheet even.  The debt-to-asset ratio, $h_i(T)/a_i(0):=f_i(0)=f_i \in
(0,1)$ captures the different capital structure between banks and $1/f_{i}$ can be seen as an approximate measure of the credit
quality of a bank.

The objective of a risk-averse bank is to increase its asset value per
unit of risk. The latter results from investing in a number
non-divisible risky projects $l \in \{1,2,3,...,N\}$, i.e. activities
related to the real-economy such as loans to firms and households,
infrastructure projects, or real-estate investments. Each of these
projects has a certain price value per unit, $\nu_{l}(t)$ at time
$t$. If this price has decreased compared to the initial price
$\nu_{l}(0)$, banks face a loss, and they gain in the opposite case.
% Once selected the number $n_{i}\leq N$ of projects to hold in the
% portfolio, each banks $i$ will keep it constant in time.

If at time $t=0$ bank $i$ acquires $x_{il}(0)$ units of project $l$ at a price $\nu_{l}(0)$, its asset is given by:
\begin{equation}
  \label{eq:asset}
  a_i(0) :=  \sum_{l=1}^{n_i} x_{il}(0) v_l (0).
\end{equation}
The investment position of bank $i$ at $t=0$ is denoted by the vector $\mathbf{x}_i(0):=\left\{x_{i1}(0),...,x_{il}(0)\right\}$ of risky
projects.  In this paper, we assume that bank $i$ invests the same proportion of its assets in each of the projects,
i.e. $a_{i}(t)/n_{i}$=const. for all $t\geq0$, which implies that Eq. (\ref{eq:asset}) simplifies to $a_{i}(0)=n_{i}\ x_{il}(0)
\nu_l (0)$, which holds for every project $l$ bank $i$ invested in.  Consequently, each bank initially has a different leverage
ratio $f_{i}$ and a different diversification expressed by the \emph{number} $n_{i}$ of projects to invest:
\begin{equation}
\label{leverage}
  \frac{1}{f_{i}}= \frac{n_{i}\ x_{il}(0) v_l (0)}{h_{i}(T)}
\end{equation}
We have assumed here that diversification is not with respect to the
\emph{type} of projects chosen, but only with respect to their
number. That means that projects, from this perspective, are
indistinguishable and uncorrelated, so we also assume that they have
the same initial price $\nu_{l}(0)=\nu_{0}$.

The balance-sheet identity (\ref{eq:36}) has to be guaranteed at any time $t$, which implies $a_{i}(t)=n_{i}\ x_{il}(t)
\nu_{l}(t)$ for the asset side to hold true. Since the fraction of assets allocated to each project is kept constant over time,
bank $i$ has to adjust the number of units $x_{il} (t)$ in its portfolio with respect to the price changes $\nu_l (t)$.
% According to a
% rebalancing strategy they bring their portfolio -- that has
% deviated away from the original equally-weighted allocation -- back into
% line, by selling over-weighted projects to purchase under-weighted
% ones. 
In practice, at each trading date, the asset allocation is revised
such that, after rebalancing, the amount invested in each of the
projects is again $a_i(t)/n_{i}$.\footnote{The theory of rebalanced
portfolios is developed in \cite{kelly1956new} and
\cite{mossin1968optimal}, among the others.} That is, bank $i$ holds the
investment position $\mathbf{x}_i(t)$ during the period $[t - dt, t]$
and liquidates it at the prevailing prices at time $t$.
Simultaneously bank $i$ sets up the new portfolio
$\mathbf{x}_i(t+dt)$.  Therefore, $\mathbf{x}_i$ represents a
``contrarian'' asset allocation strategy because the portfolio is
rebalanced by selling ``winners'' and buying ``losers'' at any time
$t$ according to the formula
\begin{equation}\label{eq:1}
%x_{il}(t+dt)=  x_{il}(t) \frac{\nu_l(t)}{\nu_l(t+dt)}
  x_{il}(t+dt)=  \frac{1}{\nu_l(t+dt)} \frac{1}{n_i} \sum_{k=1}^{n_i} x_{ik}(t)\nu_k(t+dt) 
\end{equation} 
for each project $l$ in $\mathbf{x}_i(t)$.
% This implies that they
% adjust the number of units $x_{il}(t)$ in their portfolio with respect
% to the price changes of $\nu_{l}(t)$. If the price goes down, they buy
% more units (with the hope that the price goes up, later) and the other
% way round, which is commonly known as rebalanced strategy \cite{???}.
As in 
\citet{Merton1974PricingofCorporate}, the price dynamics is given by a Geometric Brownian Motion
of the form
\begin{equation}\label{eq:8}
  \frac{d \nu_l}{\nu_{l}(t)}= \mu \, dt + \sigma\,  d\tilde{B}_l(t) \:,
  \quad l=1,...,N
\end{equation}
The random shocks $\tilde{B}_l(t)$ follow a standard Brownian motion defined on a complete filtered probability space
($\Omega;\mathcal{F}; \{\mathcal{F}_t\};\mathbb{P}$), with $\mathcal{F}_t = \sigma\{\tilde{B}(s): s \leq t \}$ and
$\mathbb{E}(d\tilde{B}_l,d\tilde{B}_y)=0$ for all pairs $l,y$ in the set of risky projects. The drift term $\mu$ describes the
instantaneous risk-adjusted expected growth rate, and $\sigma >0$ is the volatility of the growth rate. Both $\mu$ and $\sigma$
are the same for all projects because they are indistinguishable.  Despite projects are uncorrelated, diversification increases
the likelihood of having overlapping portfolios between bank $i$ and $j$, i.e. $\mathbb{E}(dB_i, dB_j)=\rho_{ij}dt$, where $\rho_{ij}$ is the asset correlation in the portfolios of banks $i$ and $j$.

Using the expression  $a_{i}(t)=n_{i}\ x_{il}(t) \nu_{l}(t)$, we
arrive after some transformations at the following dynamics for the
asset side of bank $i$:
\begin{equation}\label{eq:39}
\frac{da_{i}}{a_i(t)}= \mu dt + \frac{\sigma}{\sqrt{n_i}}\, dB_i(t) \:,
\quad \textnormal{where}\; dB_i \sim N(0, dt).
%a_{i}(t)& = a_{i} \ge 0\nonumber
\end{equation}
% Random shocks follow a normal distribution with a variance $dt$. 

For each bank $i$ the dynamics is then as follows: initially it selects a certain \emph{strategy}, defined as $(f_{i},n_{i})$,
which contains the chosen leverage $f_i$ and the number $n_i \leq N$ of risky projects, in each of which it invests the same
proportion of assets.  The bank holds the selected portfolio and locks the proportion of assets allocated to each project until
time $T$. Given the stochastic nature of the price dynamics and the risk associated with this, the bank has to estimate the
optimal combination of leveraging
%\footnote{In this case, the bank benefits from higher returns on equity. No tax-shield  effects.} 
and diversification across the set of $N$ risky projects. This problem will be addressed in the
next section.

% Consider an economy composed of two banks and a frictionless market of $N$
% non-divisible risky projects, where $v_l(t)$ is the price value per unit of project $l
% \in \{1,2,3,...,N\}$ at time $t>0$.  For simplicity, we assume that
% the projects 

% Moreover, they provide no dividend payments and their

% We will focus later on the case
% where $F$ = 2, and $N$=10 but we state the problem for the general $F$ institutions and $N$ risky assets case.

% The diversification level is: (1) complete when both banks diversify across the entire security market (i.e., $n_1=N$ and
% $n_2=N$); (2) ``almost'' complete when both banks diversify across almost
% the entire security market (i.e., $n_1=N-1$ and $n_2=N-1$).
 
% \begin{align}\label{eq:39}
% da_{i}(t) &= \frac{1}{n_i}\sum_{l=1}^{n_i}dv_{l}(t)
% =\frac{1}{n_i} \sum_{l=1}^{n_i}\mu_l dt +\frac{1}{n_i}
%   \sum_{l=1}^{n_i}\sigma_l dB_{l}(t) =a_{i}\mu_idt +
% a_{i}\frac{\bar{\sigma_i}}{\sqrt{n_i}}dB_i(t) \;,
% %a_{i}(t)& = a_{i} \ge 0\nonumber
% \end{align}
% where $\mu_i=\frac{1}{n_i} \sum_{l=1}^{n_i}\mu_l$,
%  $\frac{\bar{\sigma_i}}{\sqrt{n_i}}=
% \left(\frac{n_i^{-1}\sum^{n_i}_{l=1}\sigma^2_l}{n_i}\right)^{1/2}$ and $dB_i \sim N(0, dt)$.

\subsection{Individual and Systemic Default Probability}
\label{sec:prob}

Following the structural approach proposed by \citet{Merton1974PricingofCorporate}, we assume that the default occurs if the asset
value of bank $i$ at debt maturity date $T$ falls beneath the book value of its debt, i.e., $a_i(T) \leq h_i(T)$.  More formally,
given the probability space ($\Omega, \mathcal{F}, \mathbb{P}$), the default probability $\mathbb{P}D_i(f_i,n_i)$ of the bank $i$
using strategy $(f_{i},n_{i})$ is at the end of the period $T$ defined as:
\begin{align}\label{eq:40}
  \mathbb{P}D_i(f_i, n_i):=\mathbb{P}[a_i(t)\leq h_i \mid a_i(0)=a_{i0}]=
  \mathbb{P}[\mathrm{ln}(a_i(t))\leq \mathrm{ln}(h_i) \mid a_i(0)=a_{i0}]
 =\Phi_{1}(z_{i})
% \left(-\frac{\mathrm{ln}(1/f_{i})- \frac{\chi}{n_i}}
% { \sqrt{\frac{2 \chi}{n_i}}}     \right) \:,
\end{align}  
where $\Phi_{1}(z_{i})$ is the standard cumulative normal distribution function with the argument
\begin{equation}
z_{i}=-\dfrac{\ln(1/f_{i})+\mu T-\chi/n_{i}}{\sqrt{2\chi/n_{i}}}\:;\quad
\chi=\frac{\sigma^2 T}{2}
\label{eq:def}
\end{equation}
The constant $\chi$ combines the effects of asset volatility and time to maturity. %, and for the sake of simplicity, $\mu=0$.
The probability of bankruptcy depends on the distance between the current asset value $a_{i}(t)$ and the debt value $h_{i}(T)$, 
adjusted for the expected growth in asset value, $\mu$,  relative to asset volatility, $\sigma$. We verify that the default probability
is increasing in $h_i$, decreasing in $a_{i}$ and, for $a_{i} > h_i$, increasing in ${\sigma}/{\sqrt{n_i}}$, which is in line with
economic intuition.  Then, increasing asset diversification (i.e., letting $n_i \to N$) is desirable since it lowers
$\Phi_{1}(z_{i})$, \textit{ceteris paribus}.
%See Fig. \ref{fig:individual-def}.
% \begin{figure}[h]
% \begin{center}
% \includegraphics[width=0.8\textwidth]{PDi0}
% \end{center}
% \caption{\label{fig:individual-def}\small Individual default
%   probability as a function of the diversification level. Different
%   curves for different times to maturity of the liability.
%  \normalsize}
% \end{figure} 

%\subsection{Systemic Default Probability}

% The systemic default is commonly considered as an event that affect a
% considerable number of financial institutions/markets in a strong
% sense, thereby severly impairing the general well-functioning of
% important part of the financial system. 
%The
The individual default probability (\ref{eq:40}) refers to the default of a bank irrespective of the default of other
banks.\footnote{Indirect effects such as price movements caused by the default of other banks are not considered here.} However,
in order to quantify systemic risk, we have to specifically address the impact of the default of one bank on another bank. In this
paper, we extend the approach by \citet{Merton1974PricingofCorporate} by calculating this impact based on the above
considerations.  For simplicity, we consider only two banks in the first place and further assume the expected growth to be zero,
$\mu=0$.  A possible extension of the framework to $M$ banks is feasible (see e.g., \citet{cathcart2004multiple}) and is further
discussed in Section \ref{sec:analysis}. Also the systemic effects of a ``booming'' market, $\mu >0$, compared to those of a
``crashing'' market, $\mu<0$, are discussed later.

 Systemic default occurs whenever banks $1$ and $2$ jointly default at
the end of period $T$.  More formally,
     given the probability space ($\Omega, \mathcal{F}, \mathbb{P}$),
the systemic default probability $\mathbb{P}S(n_{1},f_1; n_{2}, f_2)$ of the banking system consisting of two banks $i\in \{1,2\}$ is defined as:
\begin{align}\label{eq:41}
  \mathbb{P}S(n_{1}, f_1; n_{2}, f_2):=\Phi_2 \left(z_{1}, \: z_{2}, 
% - \frac{  \mathrm{ln}(1/f_1)
% - \frac{\chi}{n_1}}
% { \sqrt{\frac{2 \chi}{n_1}}} \:, -\frac{
% \mathrm{ln}(1/f_2)- \frac{\chi}{n_2}}
% { \sqrt{\frac{2 \chi}{n_2}}} 
\: \rho_{1,2} \right)
\end{align} 
$\Phi_2$ is the standard bivariate cumulative normal distribution
  function and $\rho_{1,2}$ is the asset correlation between bank $1$ and
  $2$, which results from the fact that both may have invested into the same risky project.

The calculation of Eq. \ref{eq:41} implies that we have to integrate the underlying density function 
$g(z_{1},z_{2},\rho_{1,2})$ over a two-dimensional grid $(z_{1},z_{2})$, to obtain the cumulative function $\Phi_{2}$. This  is generally not simple, especially when the
asset correlation $\rho_{1,2}$ is non-constant. Various tables exist
if $g$ is a bivariate normal distribution
\citep{Pearson1931, Owen1962}. Because this does not apply to our case, we use a method proposed by \citet{Jantaravareerat}, which allows us to calculate $\Phi_{2}$ 
as follows: 
\begin{enumerate}

\item Calculate the joint probability density function
  $g(z_{1},z_{2},\rho_{1,2})$ defined as:
\begin{equation}
\label{joint-pdf}
g(z_{1},z_{2},\rho_{1,2}) = \dfrac{1}{2\pi \sqrt{1-\rho_{1,2}^{2}}}
                 \exp{\left\{-\dfrac{1}{2(1-\rho_{1,2}^{2})}\right\}}
                 \left(z_{1}^{2}-2\rho_{1,2} z_{1}z_{2}+z_{2}^{2}\right)
\end{equation}
where the $z_{i}$ are given in 
Eq. (\ref{eq:def}). This returns the joint probability density for a
  square grid with a side length, $N$, spanned by the range of $z_{1}$
  and $z_{2}$.  

\item Map the sorted values of $z_{1}$, $z_{2}$ to the index sets $I_{z_{1}}$ and $I_{z_{2}}$. I.e, the elements of $I_{z_{1}}$
  and $I_{z_{2}}$ are $y_{1}$ and $y_{2}$ that range from $[1,1]$ to $[N,N]$. Hence, after calculating the density
  $g(z_{1},z_{2},\rho_{1,2})$, we map it to $g[y_{1},y_{2}]$\footnote{Square brackets shall denote that we use the
    index instead of the actual values as coordinates}.  by using the row and column coordinates $[y_{1},y_{2}]$.

\item Calculate the mean density $d[y_{1},y_{2}]$ at each cell as the mean of itself and its three immediate neighbors.
\begin{align}
\label{mean-density}
d[y_{1},y_{2}] =\dfrac{1}{4} \Big( g[y_{1},y_{2}]+g[y_{1}+1,y_{2}]
%\nonumber \\ &
+
                g[y_{1},y_{2}+1]+g[y_{1}+1,y_{2}+1] \Big)
\end{align}

\item Calculate the volume of each cell, $P[y_{1},y_{2}]$ by
  multiplying the average density by the area, to accommodate for the fact that the indexed cells were originally of unequal size:
\begin{equation}
\label{volume}
P[y_{1},y_{2}] = d[y_{1},y_{2}]\, (z_{2}-z_{1})^{2}
\end{equation}
\item Calculate the joint distribution function, $\Phi_{2}\left[y_{1}, y_{2} \right]$ for each cell
  $[y_{1},y_{2}]$ by summing up the volumes of all cells up to this point, 
\begin{equation}
\label{cdf}
\Phi_{2}[y_{1},y_{2}] = \displaystyle \sum_{j_{1}=1}^{y_{1}} \sum_{j_{2}=1}^{y_{2}}P[j_{1},j_{2}]
\end{equation}
\item Map $\Phi_{2}[y_{1},y_{2}]$ back to $\Phi_2 \left(z_{1}, z_{2}, \rho_{1,2} \right)$ by using the index sets $I_{z_{1}}$, $I_{z_{2}}$
\end{enumerate}
In the following section we analyze Eq. (\ref{cdf}), to quantify the impact of the interplay between diversification
$(n_{1},n_{2})$ and leverage $(f_{1},f_{2})$ on the systemic default probability. Instead of the general case, where both banks
follow different strategies $(f_{i},n_{i})$, we restrict our calculation to the homogeneous case, i.e. we assume the same leverage
and diversification for both banks, i.e.  $f_1=f_2=f$ and $n_{1}=n_{2}=n$.  In Section \ref{sec:analysis}, we will also comment on
the more general case, to clarify that the homogeneous assumption is not a severe restriction.  For the homogeneous case, the
asset correlation becomes a constant expressed as
\begin{equation}
\rho_{1,2}=\frac{n^2}{N}\frac{\sigma^2}{n^2}\left/\frac{\sigma^2}{n}\right.= \frac{n}{N}
\label{eq:corr}
\end{equation}
This reduces the problem to the discussion of $\Phi_{2}(z,\rho)$ with $z(f,n,\chi)$ given by  Eq. (\ref{eq:def}). 

\section{Analysis}
\label{sec:analysis}
% \begin{figure}[htbp]
% \centering
% \includegraphics[width=4.in]{joint_cdf-single-new}
% \caption{\label{fig:individ-default-prob} Effects of market risk, degree of
%   diversification, and leverage on the individual default
%   probability $\Phi_{1}$ (color). Parameters: $f \in \{0.1, 0.3, 0.5,
%   0.7, 0.9\}$, $n \in
%   [1,40]$, $\chi \in [0.001,9]$.
% \normalsize}
% \end{figure}

\begin{figure}[htbp]
{\includegraphics[width=3.in]{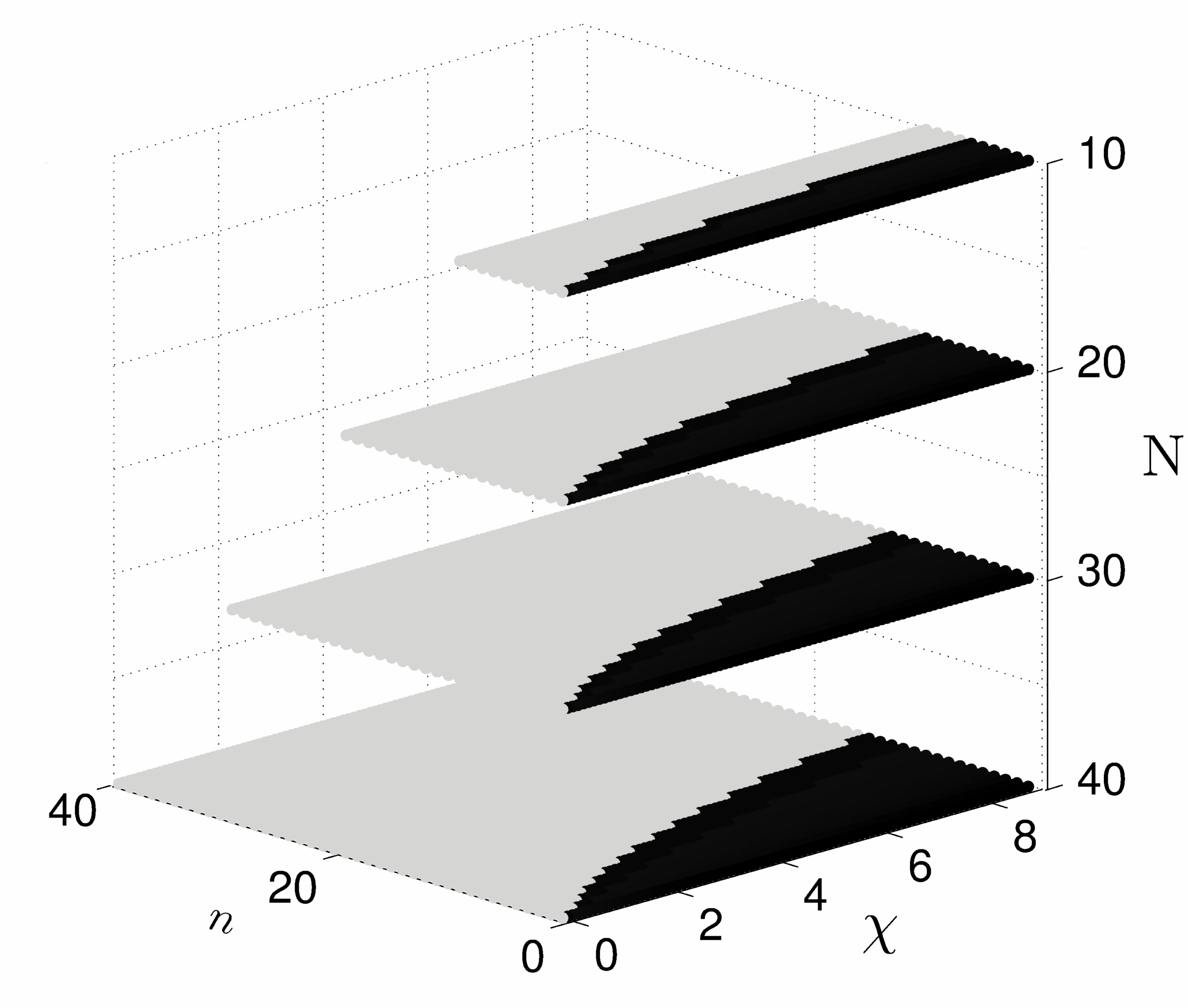}}\hfill
{\includegraphics[width=3.in]{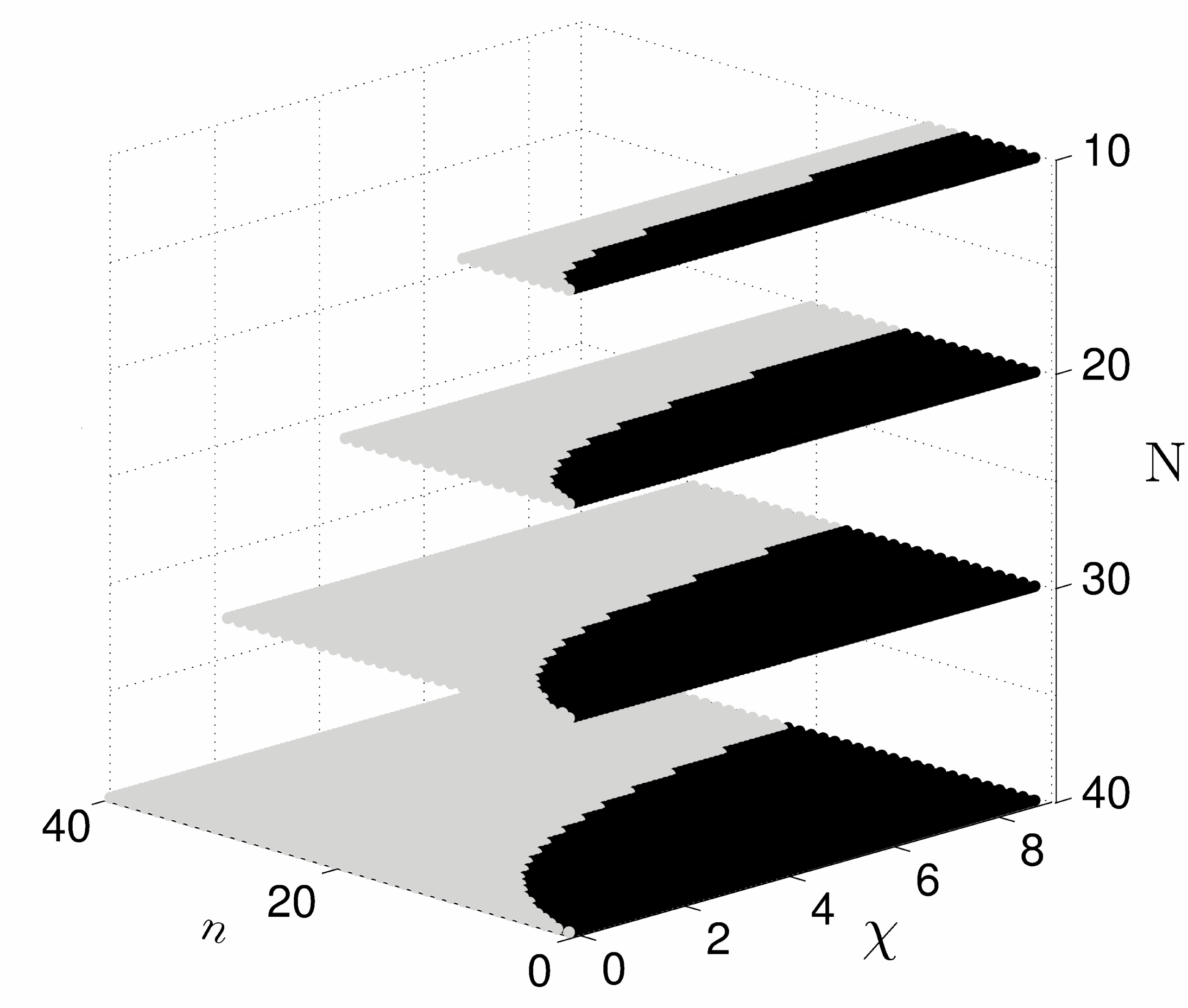}}
\caption{\label{fig:2} Effects of market size, degree of
  diversification, and leverage on the systemic default
  probability. The two plots show the difference
  $\Delta \Phi_{2} (f_{a},f_{n})$. \textbf{Left}: $\{f_{n},f_{a}\}$
  =$\{0.10,0.25\}$ and $\Delta f=0.15$. \textbf{
    Right}:$\{f_{n},f_{a}\}$ = $\{0.25,0.50\}$ and $\Delta f=0.25$. Each plane corresponds to a
  particular market size, $N$. Black
  color indicates positive values of $\Delta \Phi_{2}$, whereas gray color indicates
  zero. Other parameters: $N \in \{10,20,30,40\}$, $n \in
  [1,N]$, $\chi \in [0.001,9]$.
\normalsize}
\end{figure}

The emphasis of this section is to quantify the impact of banks'
strategy $(f,n)$, in the homogeneous case, on the \emph{systemic}
default probability $\Phi_{2}\left(z(f,n,\chi),n/N\right)$, given some
properties of the market, $\chi$. In detail, we assess the benefits of
diversification to eliminate the portion of systemic risk that is
caused by excessive leverage.  The latter is defined as the
difference $\Delta f:= f_a -f_n$, where $f_{n}$ refers to the
``normal'' market level of leverage and $f_{a}>f_{n}$ to the
``abnormal'' level.
% Let us first take a look at how the \emph{individual} default
% probability $\Phi_{1}$ depends on these parameters (see
% Fig.\ref{fig:individ-default-prob}). In accordance with intuition, if
% the volatility of the market $\sigma^{2}$ and/or the leverage $f$
% increases, the individual default probability also increases. 
% If the
% diversification $n$ increases, this probability decreases
% monotonously, which leads to the assumption that maximum
% diversification (i.e., $n \to N$)
% is always better. 
% These insights should now be contrasted with the observations from the
% \emph{systemic} default probability $\Phi_{2}$, which in the
% homogeneous case is a function
% $\Phi_{2}\left(z(f,n,\chi),n/N\right)$. 
We then calculate the difference $\Delta\Phi_{2}(f_{a},f_{n}) :=
\Phi_{2}(f_{a})-\Phi_{2}(f_{n})$ (all other parameters
fixed). $\Delta\Phi_{2}$ gives us the increase in systemic risk
associated with excessive leverage, i.e., it can be only positive
(increasing risk) or zero (no impact of the leverage), but not
negative because increasing leverage should not result in decreasing
risk.  In Figure \ref{fig:2} we plot $\Delta \Phi_{2}$ over a
parameter sweep of the market size, $N$, the market risk, $\chi$, and
the degree of diversification, $n$. 
Figure \ref{fig:2} demonstrates the
impact of the parameters $N$, $n$, and $\chi$ on the systemic risk for
two different sets of leverage values: $\{f_{n},f_{a}\}$
=$\{0.10,0.25\}$ with $ \Delta f$=$0.15$ and $\{f_{n},f_{a}\}$ =
$\{0.25,0.50\}$ with $\Delta f$=$0.25$. In the graph, we only distinguish
between zero\footnote{We talk about small positive values of
  $-10^{-6}$ or smaller, which are approximately zero, compared to an
  order of $-10$ for the black area.}  (gray) and positive (black)
values of $\Delta\Phi_{2}$. From this, we can draw the
following conclusions.

\paragraph{Diversification:} For all other values constant, there exists a diversification level $n^{\star}(f_{n},f_{a},N,\chi)$
beyond which excessive leverage does not result in an increasing systemic default probability $\Delta\Phi_{2}$. Remarkably,
$n^{\star}<<N $ (see the lines that separate the black from the gray areas in Fig.~\ref{fig:2}).
% For all other values constant, increasing $n$ leads to a critical
% value $n^{\star}(f_{a},f_{b},N,\chi)$ above which further
% diversification will not lead to a further decrease in systemic
% risk. 
In other terms, conditional on the market size $N$ and market properties
$\chi$, $n^{\star}$ represents the \textit{minimum} number of risky projects
each bank has to hold in order to prevent the increase of  systemic risk from  excessive leverage.
As Figure
\ref{fig:2} indicates, $n^{\star}$ increases monotonously with market
size $N$, i.e., a bigger market requires more
diversification. Moreover, we can show that  $n^{\star}$ increases for ``crashing'' markets,
$\mu <0$,  and decreases for ``booming'' markets, $\mu >0$. 
This is in line with economic intuition, i.e. 
diversification works well only in a booming market, \citep{tasca2011diversification}.
The specific dependence $n^{\star}(f_{n},f_{a},N,\chi)$ can be
obtained only numerically.  Table \ref{div_benefits} gives some
numbers for the minimum value, in accordance with Figure \ref{fig:2}.

% To appreciate these refined calculation based on the systemic default probability, $\Phi_{2}$, in  Table \ref{div_benefits} we also present some estimations about a minimum diversification based on a much simpler calculation.  Let us assume that the unavoidable market risk depends on diversification as $\sigma^{2}/n$, according to Eq. (\ref{eq:39}). Hence, a fraction $1/n$ of all projects will still be at risk, while a fraction $1-1/n=(n-1)/n$ of all projects could be possibly secured. This value has to be compared with the risk at maximum level of diversification, which results if all $N$ projects in the market are involved. Hence, we define the possible security level $\alpha$, i.e. the percentage of diversifiable risk, as
% \begin{equation}
%   \label{eq:alpha}
%   \alpha=\frac{1-1/n}{1-1/N}=\frac{N}{n}\frac{n-1}{N-1}
% \end{equation}
% Ideally, $\alpha$ should be above the 95\% level. For a given market
% size $N$, $\alpha$ implicitely defines the number $\bar{n}(\alpha)$ of
% projects to be chosen for a given security level, which is presented
% in Table \ref{div_benefits}. If we compare their values with the ones
% for $n^{\star}(\chi)$, which basically gives a similar information, we
% see that the naive estimation underlying Eq. (\ref{eq:alpha}) can be
% at best regarded only as an upper boundary. In fact, it overestimates
% the number of projects needed to minimumly diversify away the increase of systemic
% risk after the leverage shifted from $f_a$ to $f_b$. That is, the
% refined calculation based on the systemic default probability $\Phi_{2}$ 
% leads to much lower numbers.
\begin{table}[htbp]
 \begin{center} 
     \begin{tabular}{c|ccc|ccc}
\hline \hline
&  & $\{f_{a},f_{b}\}=\{0.1,0.25\}$  &  &
 & $\{f_{a},f_{b}\}=\{0.25,0.5\}$ & \\
       $N$ & $n^{\star}(1.6)$ & $n^{\star}(5.1)$  & $n^{\star}(8.9)$  &
$n^{\star}(1.6)$ & $n^{\star}(5.1)$ & $n^{\star}(8.9)$
\\\hline
10 & 3 & 5 & 6     & 5 & 6 & 7 \\
20 &  4&  8& 10    & 8 & 11 & 12  \\
30 &  5& 10 & 13   & 10 & 15 & 17  \\
40 &  5&  11& 15   & 12 & 18 & 22  \\
\hline \hline
      \end{tabular}
      \caption{Minimum level of diversification  $n^{\star}(\chi)$ to eliminate the increase  of systemic
        risk due to excessive leverage. \textbf{Left}:
        $\{f_{a},f_{b}\}$=$\{0.1,0.25\}$ and $\Delta
        f=0.15$. \textbf{Right}: $\{f_{a},f_{b}\}$=$\{0.25,0.5\}$ and
        $\Delta f=0.25$.  Other parameters: $N \in \{10,20,30,40\}$,
        $\chi \in \{1.6,5.1, 8.9\}$. The numbers for $n$ are rounded to the
        nearest integers.}
\label{div_benefits}
 \end{center}
\end{table}

\paragraph{Market size:}
We can observe both from Figure \ref{fig:2} and from Table
\ref{div_benefits} that an increasing market size $N$ results in an
increasing number of projects to be chosen and hold for optimal
diversification. Hence, larger markets bear more risk which has to be
compensated. Indeed, despite the fact that broader markets in which
banks can choose from a bigger pool of projects allow
for better diversification possibilities, they also reduce the
effectiveness of diversification. That is because the boundary $n^\star$
increases with the market size. In this context, systemic risk can be
hardly contained through diversification.
It could be more interesting, however, to discuss the
\emph{relative risk}, i.e. the ratio of the black area (where systemic
risk has increased with leverage increase) over the gray area (where
systemic risk is independent of leverage and diversification). It
appears that the relative risk \emph{decreases} with increasing market
size. I.e., for larger markets it becomes more likely to invest in
projects such that banks 1 and 2 become independent enough with
respect to their diversification strategy. For $n$ randomly chosen
projects, a larger market reduces the risk of systemic default,
provided that $n$ is above the critical number $n^{\star}$.

\paragraph{Leverage and market risk:} Comparing both sides of Figure \ref{fig:2}, we observe that the black area increases if the
excessive leverage $\Delta f$ increases. This also holds for increasing absolute leverage values (not shown). However, it is worth to
note that the impact of leverage is not as severe as one might expect. Its influence becomes noticeable mainly for higher values
of $\chi$, i.e. for extreme market risks. Instead, in less volatile markets, i.e. for moderate market risks, the relative impact
of the leverage decreases.

% The main message drawn from the above consideration is the existence
% of a critical boundary that distinguishes diversification strategies
% with \emph{increasing} systemic risk from those where changes in
% diversification, market size or leverage do not further impact
% systemic risk. The region of critical strategies is smaller than a
% naive diversification approach would suggest, but far from being
% negigible.

The main message drawn from the above consideration is the emergence
of two market regimes: the \textit{safe} regime in which excessive
leverage in the %
market does not increase systemic risk (gray areas) and the
\textit{risky} regime in which excessive leverage in the market does
increase systemic risk (black areas).  There are two complementary
ways to move from the risky to the safe regime. First, for a given
market condition $\chi$, banks should increase their level of
diversification at least up to $n^{\star}$ (outward movement along the
$n$ axis in Fig.~\ref{fig:2}). Second, for a given diversification
level $n$, the market should become less risky (inward movement along
the $\chi$ axis in Fig.~\ref{fig:2}).

Our results already allow us to discuss changes in the basic
assumptions that underlie the \cite{Merton1974PricingofCorporate}
model. So far, we have assumed that there are no correlations between
projects. If these are taken into account, the boundary $n^\star$
should rise dramatically to a level that may be difficult to achieve
by some market participants, especially if the market size is very
big. This effect should even further increase if we take into account
the costs for monitoring and transaction of the projects. Such costs
are expected to increase with $n$ and thus also lead to an increase in
$n^\star$. Therefore, the estimation given in this paper should be
considered as the minimum value. At the same time, this is also the
optimal diversification level because it warrants to be in the safe
regime at the minimal cost.
% deterioration of the credit quality of the banking system)

Our model could be extended to the more general case of $M$ banks, which may be heterogeneous in terms of leverage and
diversification strategies $(f_{i},n_{i})$. This would lead to an increasing complexity of the analysis. However the general
findings of the paper would qualitatively remain the same because such generalizations do not increase the degree of freedom for
the banks, i.e. their set of strategies. At this point, we can already argue that the impact of the heterogeneity in strategies is
as follows: If some banks choose a high level of leverage, this increases $n^{\star}$. This in turn forces those other banks with
lower leverage to choose a larger diversification level, to compensate for this, i.e. it creates negative externalities in the
system.

\section{Conclusions}
\label{sec:conclusions}
% The aim of this paper was to address the impact of failing agents on the default of other agents. 
% In order to describe the stability of an individual agent, we used the framework introduced by \citet{Merton1974PricingofCorporate}. It allows us to consider two main factors XXXX

%%%%%%%%%%

% In the ongoing process of innovation, creation and
% integration of financial markets, the study of the systemic effects
% of individual corporate structures and strategies has raised at the
% top of the policy agenda \citep{haldane2009rethinking}.

Financial institutions such as banks use different strategies to run
their businesses. In this paper, we consider two of them, namely
leverage and diversification. The first one allows banks to increase
their expected return on equity based on debt financed
investments. The second one is used to mitigate losses from such
investments. Hence, both strategies have a different impact on the
default probability of a bank: leverage amplifies the default risk,
while diversification reduces it. In a finite market banks cannot
choose completely independent diversification strategies. Instead,
their portfolios overlap with diversification, which correlates the
default probability of individual banks.

In this paper, we are interested in the impact of these strategies on
the systemic default probability, i.e. the probability of joint
defaults. In particular, we answer the question if and how an increase
of individual risk, due to excessive leverage, leads to an increase of
the systemic default probability, and if and how this can be mitigated
by an optimal diversification level. As the main result, we determine
the boundary between two market regimes: a safe regime in which excess
leverage \emph{can} be mitigated by better diversification, thus
\emph{not} resulting in an increase of systemic risk, and a risky
regime in which this is not possible resulting in increased systemic risk.

Our findings contribute to the ongoing discussions of the optimal
regulatory framework for financial markets. The question, whether
benefits from diversification can offset the systemic risk, due to
excessive leverage, can be answered by means of our distinction
between risky and safe conditions. This also extends to the question
whether capital requirements should be lowered for institutions with
more diversified investments. Here, we show that a critical level of
diversification exists which depends on the market size and the market
conditions and thus needs to be monitored to prevent a transitions
toward the risky regime.
Such insights can be used to improve macro-prudential regulations,
such as the current Basel III risk-based framework, which is portfolio
invariant, i.e., does not consider the role for diversification. As a
conclusion from our results, portfolio invariance leads to a
significantly higher maximum leverage ratio allowed for banks in the
current regulatory framework, i.e. it can result in a much riskier
financial system.

\end{document}